\documentclass[11pt,a4paper]{article}

\usepackage{enumerate,amsmath,amsthm,amssymb,latexsym,
amsfonts,amscd}
\usepackage[all]{xy}

\title{The Einstein equation and \\
the energy density of the gravitational field \\}

\author{
Maurice J. Dupr\'e \\ Department of Mathematics, Tulane University\\
New Orleans, LA 70118 USA\\ mdupre@tulane.edu \\ \\}

\theoremstyle{plain}

\theoremstyle{definition}

\numberwithin{equation}{section}

\newcommand{\R}{\mathcal R}

\newcommand{\bR}{\mathbb{R}}

\newcommand{\lra}{\longrightarrow}

\newcommand{\del}{\partial}

\newcommand{\med}{\medbreak}

\begin{document}

\maketitle

\begin{abstract}
We give a derivation of the Einstein equation for gravity which
employs a definition of the local energy density of the
gravitational field as a symmetric second rank tensor whose value
for each observer gives the trace of the spatial part of the
energy-stress tensor as seen by that observer. We give a physical
motivation for this choice using light pressure.
\end{abstract}


\med \textbf{Mathematics Subject Classification (2000)} : 83C05,
83C40, 83C99.

\med \textbf{Keywords} : Gravity, general relativity, Einstein
equation, energy density.

\section{INTRODUCTION}

\med

Since Einstein's and Hilbert's original "derivations" of the
Einstein equation for gravity in classical general relativity
(CGR), there have appeared too many to list.  The many different
types of derivations are summarized in \cite{SACHSWU}. In fact,
all these subsequent derivations as well as Hilbert's original
derivation contrast markedly from Einstein's original derivation
in that they appeal to some abstract mathematical principal which
though desirable, is usually not justifiable beyond mere desire.
For instance, one of the most popular derivations simply modifies
one side of the equation to make it have zero divergence on
grounds that physical considerations make the other side, the
energy-stress tensor, have zero divergence. On the other hand, in
Einstein's original derivation, \cite{EINSTEIN2}, we see the
realization that mathematically the Ricci tensor should be
proportional to the source which should be the total energy
density due to both the energy-stress tensor as well as the
gravitational field itself. However, in \cite{EINSTEIN2}, Einstein
was not able to arrive at a tensor expression for the energy
density of the gravitational field.  Instead, he arrived at a
doubly indexed quantity which he admitted was not a tensor, but
rather a pseudo-tensor defined in terms of the the connection
coefficients and which served to give the energy of the
gravitational field for purposes of deriving the equation. As the
arguments in \cite{EINSTEIN1} leading up to the development of CGR
show, Einstein was clearly thinking of the energy of the
gravitational field in a Newtonian way, since in particular, the
connection coefficients are the generalized gravitational forces
from the Newtonian viewpoint. Moreover, in \cite{EINSTEIN2},
Einstein was very clear that his equation was using the energy
density of the gravitational field in addition to the
energy-stress tensor as the total source of gravitiy. In fact,
subsequent attempts to mathematically characterize the energy of
the gravitational field have all basically clung to the Newtonian
framework which makes the energy of the gravitational field a
function of a non-local arrangement of masses and energies.  So
much so, that these views are now taken for granted to the point
that in \cite{MTW} we have the claim of the impossibility of
existence of a local energy density tensor for the gravitational
field (see also
\cite{WALD},\cite{HAWKINGELLIS},\cite{FRANKEL},\cite{CW},\cite{SZ}).
This attitude clearly persists to the present as expressed, for
instance, in chapter 3 of \cite{SZ}. The result has been a
profusion of mathematically inspired notions of quasi-local mass,
which all have their advantages and drawbacks as discussed in
\cite{SZ} and \cite{YAU3}, along with extremely involved analysis
required to arrive at their basic properties. Fortunately, I was
unaware of these problems when I set out to find the energy
density of the gravitational field in order to derive Einstein's
equation. By contrast, we can realize the physical representation
of the local energy density of the gravitational field by thinking
of the physics of the gravitational field in relativistic terms.
When we do, we see that the resulting second rank tensor added to
the energy-stress tensor can serve as the total energy density
source for the gravitational field which then must be proportional
to the Ricci tensor.  The result in particular immediately gives
the Einstein equation. The trick is to adopt a truly relativistic
attitude towards what disturbance of the gravitational field
entails, and using laser light pressure as a standard, to relate
the disturbance back to the energy-stress tensor itself. In
particular, we find that the divergence of the
energy-momentum-stress tensor due to matter and fields other than
gravity must be zero as a consequence of our derivation.

\med

\section{THE RICCI TENSOR AND DIVERGENCE}

We assume that our spacetime is a 4-manifold $M$ equipped with a
Lorentz metric tensor $g,$ with signature $(-,+,+,+),$ and we
denote by $\nabla$ the resulting Kozul connection or covariant
differentiation operator on $M.$ We use $T_mM$ to denote the
tangent space to $M$ at $m \in M.$ It is convenient in this
setting to refer to $u \in T_mM$ as a unit vector to mean merely
$|g(u,u)|=1.$ We have then the Riemann curvature operator, $\R,$
given by

\begin{equation}\label{curvatureoperator}
\R(u,v)=[\nabla_u,\nabla_v]-\nabla_{[u,v]},
\end{equation}
where $u$ and $v$ are any tangent vector fields on $M.$ We note
that $\R(u,v)$ actually defines a vector bundle map of the tangent
bundle $TM$ to itself covering the identity map of $M,$ and it as
well then determines the Riemann curvature tensor, $R,$ of fourth
rank, which means that $\R$ is itself a linear transformation
valued tensor field on $M.$ One of our main concerns is the
certain contraction of $R$ known as the Ricci tensor, $Ric.$ In
fact in any frame at $m \in M$ with basis $e_{\alpha}$ for $T_mM$
and dual basis $\omega^{\alpha},$ We have, using the summation
convention,

\begin{equation}\label{ricci1}
Ric(u,v)=\omega^{\alpha}(\R(e_{\alpha},u)v),~~u,v \in T_mM.
\end{equation}
Among the many symmetries of the Riemann curvature tensor is the
fact that $Ric$ is a symmetric tensor.

In order to see how the Ricci tensor enters into the theory of
gravitity, we should recall the equation of geodesic deviation. If
$[-a,a]$ and $[-b,b]$ is a pair of intervals in $\bR,$ then a
Jacobi field is a smooth map $J:[-a,a] \times [-b,b] \lra M$ such
that for each fixed $s \in [-a,a]$ the map $J_s ; [-b,b] \lra M,$
given by $J_s(t)=J(s,t),$ is a unit speed geodesic in $M.$  We can
then form local vector fields $e,u$ on an open neighborhood of the
image of $J$ in $M,$ denoted $Im~J,$ so that

\begin{equation}
e(J(s,t))=\del_sJ(s,t),~~~u(J(s,t))=\del_tJ(s,t).
\end{equation}
Thus we must have $[e,u]=0$ and $\nabla_uu=0,$ on $Im~J,$ so we
find

\begin{equation}\label{Jacobi1}
\R(e,u)u=-\nabla_u\nabla_eu.
\end{equation}
We will call $e$ in this situation a tangent Jacobi field along
$J_0.$ In fact, given $m$ a point on $J_0$ and any unit vector
$e_m \in T_m$ which is orthogonal to $u(m),$ we can arrange that
$e(m)=e_m.$ Since our connection is assumed to be the unique
torsion free metric connection, we have
$[e,u]=\nabla_eu-\nabla_ue,$ so the condition that $[e,u]=0$ gives
$\nabla_eu=\nabla_ue$ in our present case. In view of
(\ref{Jacobi1}), we then find the equation of geodesic deviation
on $Im~J.$

\begin{equation}\label{Jacobi2}
\R(e,u)u+\nabla_u\nabla_ue=0.
\end{equation}
Now the term $\nabla_u\nabla_ue$ should be interpreted as the rate
of change of separation acceleration in direction $e$ of
infintesimally separated geodesics. If $\delta s$ is a small
change in the parameter $s,$ then we can think of $(\delta s)e$ as
representing the separation between the geodesic $J_0$ and the
geodesic $J_{\delta s}.$ Thus $\nabla_u[(\delta s)e]$ represents
the rate of change of separation from $J_{\delta s}$ as seen by an
observer moving along $J_0.$ Then $\nabla_u\nabla_u[(\delta s)e]$
represents acceleration in separation between the observer
following $J_0$ and the geodesic $J_{\delta s}.$ Dividing by
$\delta s$ then gives the rate of change in the direction $e$ of
the separation acceleration.  This means that
$g(e,\nabla_u\nabla_ue)=g(e,\R(e,u)u)$ is a "term" in a typical
spatial divergence calculation, in this case of the acceleration
field. That is, it is the $e-$component of the rate of change of
separation acceleration in direction $e.$ If $u$ is assumed
time-like, it follows from (\ref{ricci1}) that $Ric(u,u)$ is in
fact giving an invariant form of a spatial divergence of the
separation acceleration field for infinitesimally close geodesics
to $J_0,$ as would be seen by an observer following along $J_0.$
To be clearer about this, an observer following $J_0$ could view
the separation of nearby geodesics as a position vector field on
$u^{\perp},$ the orthogonal complement of his velocity vector, so
the change in separation would then be viewed as the spatial
velocity of nearby points (test particles suspended) in his space,
so its rate of change is the acceleration field ${\bf a}_u$ as
seen by the observer. We then have for $e \in T_mM$ that

\begin{equation}\label{acceleration field}
{\bf a}_u(e)=\nabla_u \nabla_u e,~~e \in u^{\perp} \subset TM.
\end{equation}
If $(e_1,e_2,e_3)$ is a smooth frame field along $J_0$ for the
orthogonal complement of the time-like unit vector field $u$ along
$J_0,$ which is orthonormal at $m$ on $J_0,$ then from
(\ref{ricci1}) and (\ref{acceleration field}), we have at $m,$

\begin{equation}\label{ricci2}
Ric(u,u)=-[g(e_1,\nabla_u \nabla_u e_1)+g(e_2,\nabla_u \nabla_u
e_2)+g(e_3,\nabla_u \nabla_u e_3)]=-div_u {\bf a}_u,
\end{equation}
where ${\bf a}$ denotes the spatial separation
(compare\cite{ONEIL}, 8.9, page 219 and \cite{SACHSWU}, 4.2.2,
page 114) acceleration field around $J_0,$ and $div_u {\bf a}$
denotes the spatial divergence of this spatial vector field around
$J_0.$ At this point, one might object that the observer could be
rotating which would introduce fictional acceleration into ${\bf
a}_u,$ and that is correct.  A more sophisticated analysis here
could deal with this purely mathematically (see for instance
\cite{FRANKEL}, \cite{MTW}, or \cite{SACHSWU}), but let us allow
that the observer can feel if he is rotating and just say he
restricts to cases where he is not rotating in order to carry out
his measurements. Continuing then, for a non-rotating observer,
the separation acceleration field in a geometric theory of gravity
is the essence of the gravitational field. That is, if an observer
at event $m \in M$ has velocity $u,$ with $g(u,u)=-1,$ then
according to (\ref{Jacobi2}), (\ref{acceleration field}), and
(\ref{ricci2}) we should interpret $R(u,u)$ as the negative
divergence of the gravitational "force per unit mass" field as
seen by that observer at $m \in M.$ Now, in ordinary vector
analysis, the divergence of an irrotational vector field, divided
by $4\pi,$ is its source density, and we know that the source
density for gravity is energy density. Therefore there should be a
universal positive constant $G$ so that $(1/4\pi)Ric(u,u)$ is
equal to the energy density observed by this observer multiplied
by $G,$ as gravity always acts as an attractive influence between
objects. But, thinking relativistically, if the gravitational
field itself has energy, that energy density must be included in
the source total energy density. Thus, the question now arises as
to the energy density of the gravitational field itself.

\section{THE ENERGY DENSITY OF THE GRAVITATIONAL FIELD}

In order to deal with the energy density of the gravitational
field, we must first think in terms of the basic assumption of the
geometric notion of gravity which is that "free test" particles
must follow geodesics.  If this is the case, then from the point
of view of the gravitational field itself, it is happiest when all
particles are following geodesics. In fact, we can imagine that in
a limiting sense, if "all particles" follow geodesics, then the
gravitational field is completely relaxed and contains no  energy.
It is only when we try to push a particle off of its geodesic that
we feel the reaction of the gravitational field, and notice we
feel it right at the location of the event of trying to push the
particle off of its geodesic. Thus, relativistically, we should
think of the manifestation of tension in the gravitational field
is particles not following geodesics. Now, if a particle is not
following a geodesic, then it is because it is being acted on by a
force which is not part of the gravitational field itself. Because
by definition, gravity acts only through causing particles to
follow geodesics, in the absence of "outside" forces. When a force
acts to move a particle a certain amount off of its geodesic path,
the force required to do so is proportional to the particles
inertial mass, by definition, but in essence, this says the
gripping energy of the gravitational field at the point where the
particle is located is somehow related to the mass of the
particle. Accepting this, the force density should be manifested
in pressures in various directions. That is, the energy-stress
tensor tells us the pressures as seen by any observer in various
directions, so from the energy-stress tensor itself, we should be
able to find the energy density of the gravitational field. For
instance, the pressure you feel on your bottom when sitting in a
chair is a manifestation of the energy density of the
gravitational field at those points on your chair. In a sense
then, we could say that if the surface of your chair were replaced
by an infinitesimally thin slab sitting on top of an
infinitesimally lower chair, then the mass energy of the slab
required to hold you in place divided by the volume of the slab is
a reflection of the energy density of the gravitational field.
What is the minimum mass which can take care of this job? In fact,
the material the chair is made of in some sense is a reflection of
the energy density of the gravitational field right where your
chair is located. Even primitive people have an intuitive idea of
the strength of material needed to make a chair, and thus have a
working idea of the energy density of the gravitational field. We
should therefore think of the least mass energy of material
required to make a chair as a rough measure of the energy density
of the gravitational field where the chair is to be used. More
generally, imagine an observer located at $m \in M,$ ghost-like
inside a medium with energy-stress tensor $T$ and suppose that his
velocity at $m$ is $u.$ Then exponentiating $u^{\perp} \subset
T_mM,$ the orthogonal complement of $u$ in $T_mM,$  the pressures
are given by the restriction of $T_m$ to $u^{\perp} \times
u^{\perp}.$ This is a symmetric tensor on a Euclidian space so can
be diagonalized. Notice that this does not mean $T$ itself is
diagonalizable. Thus, there is an orthonormal frame
$(e_x,e_y,e_z)$ for $u^{\perp}$ with the property that
$T(e_a,e_b)=p_a \delta^a_b,$ for $a,b \in \{x,y,z\}.$ Imagine
scooping out a tiny box in $M$ at $m$ whose edges are parallel to
these principal axes of this spatial part of the energy-stress
tensor. We can imagine putting infinitesimally thin reflecting
mirrors for walls of the box and filling the box with laser beams
reflecting back and forth in directions parallel to the edges of
the box with enough light pressure in each direction to balance
the force from outside on these reflecting walls. In a sense, we
have standardized a system to balance the pressures acting to
disturb the gravitational field, so we define the energy density
of the gravitational field as seen by our observer to be the
energy density of the light in this little box. The fact that a
photon has zero rest mass should mean that the light energy
constitutes a minimum amount of energy to accomplish this task of
balancing the gravitational energy. However, it is an elementary
problem in physics to see that the energy density of the light
along a given axis is exactly the pressure in that principal
direction. Let us review this simple argument. Assume the
coordinates are $(t,x,y,z)$ for simplicity and the box edges are
parallel to these axes with lengths $\delta x, \delta y, \delta
z,$ respectively. Assume that the laser beams parallel to say the
$x-$axis contain $N_x$ photons, each having spatial momentum
$P_x.$ In time $\delta t,$ the photons travel a distance of
$c\delta t$ and hence each such photon makes $(c \delta t)/(\delta
x)$ reflections for a change in momentum of $2P_x$ for each
reflection. Thus the total momentum transfer to the two end walls
perpendicular to the $x-$axis for the laser beams paralleling the
$x-$axis is

\begin{equation}\label{laser1}
\frac{2P_xN_xc \delta t}{\delta x}.
\end{equation}
This means that the force exerted on the two end walls is
$(2P_xN_xc)/(\delta x).$ But the total area of the two end walls
is $2\delta y \delta z,$ so the pressure on the end walls is

\begin{equation}\label{pressure x}
p_x=\frac{N_xP_xc}{V},
\end{equation}
where $V=\delta x \delta y \delta z$ is the volume of the box. But
the relativistic energy of a photon with momentum $P_x$ is $P_xc.$
Therefore, the total energy density due to the $x-$axis beams is
exactly the pressure in the $x-$direction on the walls
perpendicular to the $x-$axis. Likewise for the other two axes,
consequently we see that the energy density in the box is the sum
of the pressures that the beams are balancing, that is the trace
of the observer's spatial part of the energy-stress tensor,
$p_x+p_y+p_z.$

\section{THE ENERGY DENSITY TENSOR\\ OF THE GRAVITATIONAL FIELD}

In this section, and the next, we will be applying what we will
call the {\it symmetric tensor observer principle} (see Appendix
on Analytic Continuation). In our situation, as applied to second
rank tensors, it means that if $A$ and $B$ are both symmetric
second rank tensors at $m \in M,$ and if $A(u,u)=B(u,u)$ for all
unit time-like vectors, it follows purely mathematically, that
$A=B.$ This is the essence of the {\bf Principle of Relativity} as
applied to second rank symmetric tensors-a law (at $m$), say
$A=B,$ should be true for all observers (at $m$) and conversely,
if true for all observers (at $m$), that is if $A(u,u)=B(u,u)$ for
all time-like unit vectors $u \in T_mM,$ then it should be a law
(at $m$) that $A=B.$ We are going to obtain the Einstein equation
by simply observing that if $H$ is the symmetric tensor giving the
total energy density at $m \in M,$ then $(1/4\pi
G)Ric(u,u)=GH(u,u)$ for each vector $u \in T_mM$ such that
$g(u,u)=-1.$ First, let $c(A)$ denote the contraction of $A$ for
any second rank covariant tensor $A.$ Thus using the summation
convention, we have

\begin{equation}\label{contraction}
c(A)=g^{\alpha \beta}A_{\alpha \beta}.
\end{equation}
Now suppose that $u \in T_mM$ is the velocity of an observer at $m
\in M.$  Suppose that $T$ is the second rank covariant
energy-stress tensor.  We can define the projection tensor
$P_u:T_mM \lra T_mM$ by $P_u(w)=w+g(w,u)u,$ for any $w \in T_mM.$
Note here that we are not assuming $w$ is time-like or space-like,
it is just any vector in $T_mM.$ We easily check that $P_u$ is
linear, that $P_u \circ P_u=P_u,$ and that $P_u(T_mM)=u^{\perp}
\subset T_mM.$ For instance, if $w \in T_mM,$ then
$g(u,P_u(w))=g(u,w)+g(w,u)(-1)=0,$ so $P_u(T_mM) \subset u^{\perp}
.$ On the other hand, if $w \in u^{\perp},$ then obviously
$P_u(w)=w,$ and therefore, $P_u(T_mM)=u^{\perp}.$ Thus,
$A(v,w)=T(P_u(v),P_u(w))$ is a symmetric tensor at $m$ and
restricted to $u^{\perp},$ this tensor can be diagonalized as $g$
on $u^{\perp}$ is positive definite. That is to say, there is an
orthonormal frame $(e_1,e_2,e_3)$ for $u^{\perp}$ so that
$A(e_i,e_j)=\lambda_i \delta ^i_j.$ Thus, we see that the box
argument of the preceding section shows that the sum of these
diagonal values of $A$ must represent the energy density of the
gravitational field, namely, $a=A(e_1,e_1)+A(e_2,e_2)+A(e_3,e_3).$
Now, if we compute $c(T)$ in the frame $(u,e_1,e_2,e_3),$ we find
that $c(T)=-T(u,u)+a,$ and therefore,
$a=T(u,u)+c(T)=T(u,u)-c(T)g(u,u),$ which is to say finally that
the second rank symmetric covariant tensor $T_g,$ where

\begin{equation}\label{gravenergydensity}
T_g=T-c(T)g
\end{equation}
represents the energy density of the gravitational field at $m \in
M,$ because by the symmetric tensor observer principle, $T_g$ at
$m \in M$ is completely determined by specifying $T_g(u,u)$ for
each time-like unit vector $u \in T_mM.$

\section{THE DERIVATION OF THE EINSTEIN EQUATION}

 Now, finally, from the fact that $(1/4\pi)Ric(u,u)$ is the
 divergence of the spatial gravitational acceleration per unit
 mass, using (\ref{gravenergydensity}) our gravitation equation
 should be that for every timelike unit vector (that is every
 observer) at $m \in M,$ we have

 \begin{equation}\label{einsteinequation1}
 \frac{Ric(u,u)}{4 \pi}=G[T(u,u)+T_g(u,u)]=G[T(u,u)+T(u,u)-c(T)g(u,u)].
 \end{equation}
 Since (\ref{einsteinequation1}) is homogeneous of order 2, it
 follows that the equation is true for $u$ being any timelike
 vector, and hence by the symmetric tensor observer principle, we
 must have

 \begin{equation}\label{einsteinequation2}
 \frac{Ric}{4 \pi}=G[2T-c(T)g]=2G[T-(1/2)c(T)g].
 \end{equation}
 Let us denote the total energy-stress tensor by
 $H=T+T_g=2T-c(T)g.$  In these terms, we have simply

 \begin{equation}\label{einsteinequation21}
 Ric=4 \pi G~ H.
 \end{equation}
 Equivalently, we have

 \begin{equation}\label{einsteinequation3}
 Ric=8\pi G[T - (1/2)c(T)g]
 \end{equation}
 which is a well-known form of Einstein's equation. As $c(g)=4$
 and $c(Ric)=R,$ where as usual, $R$ is the scalar curvature, we
 find that $R=(8\pi)G[-c(T)],$ so the equation can be also written
 as $Ric=8\pi GT+(1/2)Rg,$ and this results immediately in the most
 familiar form of the Einstein equation

 \begin{equation}\label{einsteinequation4}
 E=Ric-(1/2)Rg=8\pi G~T,
 \end{equation}
where $E=Ric-(1/2)Rg$ is the Einstein tensor.
 Notice that we have not used local conservation of energy,
 $divT=0.$  Since the left side of (\ref{einsteinequation4}), the Einstein
 tensor, $E,$ is divergence free, we find $divT=0$ as a consequence of our
 derivation.

 We can observe that our derivation required the assumption that
 the pressures are positive in order for the laser light box
 argument to justify the expression on the right side of
 (\ref{einsteinequation2}) as the total energy density tensor of
 the matter fields and gravitational fields, but the development
 is so general at this point, that it seems reasonable that the
 expression (\ref{gravenergydensity}) should be regarded as the
 energy density of the gravitational field in all cases. This
 means that the total energy density is the expression $H$ on the
 right hand side of (\ref{einsteinequation21}),

 \begin{equation}\label{totalenergydensity}
 H=T+T_g=2T-c(T)g=2[T-(1/2)c(T)g].
 \end{equation}
 Reconsidering $T_g=T-c(T)g,$ it is probably more natural to think
 of the gravitational field's energy-stress tensor, $T_g$ as a
 function of the metric tensor in some way.  For this we just use
 the Einstein equation itself. Since $c(T)=(-1/8\pi G)R,$ from
 (\ref{einsteinequation4}) we immediately conclude that

 \begin{equation}\label{gravenrgydensitygeoform}
 T_g=(1/8\pi G)[Ric+(1/2)Rg]=(1/8\pi G)(E+Rg).
 \end{equation}
 From the last expression on the right, we see, as $T$ and the Einstein
 tensor, $E=Ric-(1/2)Rg$ both have zero divergence, that

 \begin{equation}\label{energydivergence}
 divH=divT_g=(1/8\pi G)div(Rg)=(1/8\pi G)dR=-d[c(T)].
 \end{equation}
 So even though the total energy and gravitational energy are not
 infinitesimally conserved, the divergence is simply proportional to the exterior
 derivative  the scaler curvature. Of
 course,  $divT_g=-d[c(T)]$ is obvious
 from the definition, (\ref{gravenergydensity}), once we accept $divT=0.$
 In particular, as $d^2=0,$ this means that

 \begin{equation}\label{exteriorderivative of grav divergence}
 d[divH]=d[divT_g]=0,
 \end{equation}
but (\ref{energydivergence}) is even better as it shows $divT_g$
is an exact 1-form on $M.$  On the other hand, the equation
(\ref{energydivergence}), when written

\begin{equation}\label{divgravenergy1}
divT_g+d[c(T)]=0
\end{equation}
has another interpretation.  In classical continuum mechanics
written in four dimensional form of space plus time, the
divergence of the energy stress tensor equals the density of
external forces.  Of course in relativity, the energy stress
tensor $T$ contains everything and there are no external forces,
as gravity is not a force.  But, we can view
(\ref{divgravenergy1}) as saying that from the point of view of
the gravitational field, the matter and fields represented by $T$
are acting on the gravitational field as an external force density
of $-d[c(T)].$  In classical continuum mechanics, the external
force density has zero time component, but relativistically such
is not the case, the force only has zero time component in the
instantaneous rest frame of the object acted on. We can therefore
view (\ref{divgravenergy1}) as saying that the divergence of the
the gravitational field's energy stress tensor is being balanced
by the rate of increase of $-c(T).$ If $p_x,p_y,p_z$ are the
principal pressures in the frame of an observer with velocity $u,$
where $g(u,u)=-1,$ then $\rho=T(u,u)$ is the energy density
observed, and $divT_g(u)$ is then the power loss density of the
gravitational field. Now $c(T)=-\rho+p_x+p_y+p_z,$ so
(\ref{divgravenergy1}) becomes

\begin{equation}\label{divgravenergy2}
divT_g(u)=D_u\rho-D_u[p_x+p_y+p_z].
\end{equation}
Thus, the observer sees the divergence of energy of the
gravitational field is exactly the rate of increase of energy
density of the matter and fields less the rate of increase of
principal pressures.  In particular, in any dust model of the
universe (pressure zero), the gravitational energy dissipation is
exactly balanced by the rate of increase of energy density of the
matter and fields. If $T$ is purely the electromagnetic stress
tensor in a region where there are only electromagnetic fields,
then $c(T)=0,$ and the gravitational energy-stress tensor has zero
divergence, so is then infinitesimally conserved.

\section{THE GRAVITATION CONSTANT $G$}

So far, we have not said anything about the determination of the
gravitation constant $G.$  To evaluate this, we merely need to
check the results of experiments with attractive "forces" between
masses. But it is much simpler to just use Newtonian gravity in an
easy example where the results should be obviously approximately
the same.  Consider an observer situated at the center of a
spherical dust cloud of uniform density $\rho,$ and calculate the
separation acceleration field using Newton's law of gravitation.
At distance $r$ from the center, but inside the cloud, the mass
acting on test particles at radial distance $r$ is simply the mass
inside that radius, $M(r),$ by spherical symmetry, as is
well-known in Newtonian gravitation. Here, we have $M(r)=(4/3)\pi
r^3 \rho.$ But Newton's Law says the acceleration of a test mass
near the center of the dust cloud is radially inward, and if $r$
is the distance from the center, then the radial component of
acceleration is given by

\begin{equation}\label{Newton1}
a_r(r)=-G_N\frac{M(r)}{r^2}=-G_N \frac{4 \pi \rho r}{3}.
\end{equation}
Here, $G_N$ is the Newtonian gravitation constant. On the other
hand, considering an angular separation of $\theta,$ the spatial
separation is $s=r\theta,$ so the relative acceleration of nearby
test particles in the $s-$ direction perpendicular to the radial
direction is therefore

\begin{equation}\label{Newton2}
a_s(r)=\theta a_r(r)=-G_N\frac{4 \pi \rho r \theta}{3}=-G_N\frac{4
\pi \rho s}{3}.
\end{equation}
Thus the rate of change of separation acceleration of nearby
radially separated test particles in the radial direction at given
$r$ is by (\ref{Newton2}),

\begin{equation}\label{Newton3}
\frac{da_r}{dr}=-G_N \frac{ 4 \pi \rho}{3},
\end{equation}
whereas in the $s$ direction we have the rate of change of
separation acceleration is

\begin{equation}\label{Newton4}
\frac{da_s}{ds}=-G_N \frac{ 4 \pi \rho}{3},
\end{equation}
the same result again.  But there are two orthogonal directions
perpendicular to the radial, so now we see that if ${\bf a}_u$
denotes the spatial acceleration field around our observer at the
center of the dust cloud, then

\begin{equation}\label{Newton5}
div_u({\bf a}_u)=-G_N 4 \pi \rho.
\end{equation}
As we are dealing with dust, the pressures are zero, so there is
no gravitational energy density, and thus $\rho$ is now the total
energy density seen by our observer.  Thus, we have by
(\ref{ricci2}), that $R(u,u)=G_N 4 \pi \rho=4 \pi G_N ~H(u,u).$
But now comparing this result with (\ref{einsteinequation21}), we
see that we must have $G=G_N.$ Notice that in our development, we
have used the symmetric tensor observer principle as a form of the
principle of general relativity to reduce everything to working
with the time component in an arbitrary frame for the tangent
space. The trick is to be able to work completely generally so
that conclusions apply to $T_{00}$ and $Ric_{00}$ no matter the
coordinate frame, which seems best expressed by using $T(u,u)$ and
$R(u,u),$  to remind us that we are dealing with an arbitrary
time-like unit vector. It is only now at the end once we have
Einstein's equation that we allow a calculation in a special frame
in order to evaluate the gravitation constant.

At this point  let us discuss for a moment the derivation of
Einstein's equation given in \cite{FRANKEL}. In effect, the
derivation of the Einstein equation given in \cite{FRANKEL} uses
the analysis of the special case of a static arrangement of mass
for a gravitating fluid drop and adds the Newtonian energy density
of the fluid drop as expressed in terms of pressure through the
requirement that its surface pressure be zero to get the time
component of the Einstein equation. Since the setup is a special
arrangement of mass, one cannot assert the symmetric tensor
observer principle, because the only observer for which the
equation works is the special observer moving with the drop.
However, one can appeal to the general covariance desire of
relativity that equations should be tensor equations valid in all
frames, from which one surmises that if you have found an equation
relating the time components in a special frame, then the other
components in that special frame should also be equal. Once you
accept the full tensor equation in any frame, then it is valid in
all frames and you next surmise that if it works for the liquid
drop, then it must work in general. But, in our present situation,
we have the full equation, and can simply go backwards through the
development in \cite{FRANKEL} to see that the time component of
the equation in the liquid drop case is Newton's law, and
therefore again conclude that our $G$ in (\ref{einsteinequation4})
is identical to the Newtonian gravitational constant. For a
treatment of linearized Einstein gravity and its Newtonian
approximation in general, on can consult \cite{MTW} or
\cite{WALD}.

At this point, we can simply choose units such that $G=1$ and we
henceforth drop this factor from the equation for simplicity.

\section{ENERGY CONDITIONS}

Since we have an expression for the total energy density $H=T +
T_g,$ we could surmise that in general it would be reasonable to
have $H(u,u) \geq 0,$ for every time-like tangent vector $u \in
TM.$ But, this is the same as requiring that $2T(u,u) \geq -c(T),$
a condition known as the {\it strong energy condition.} In
formulating the various energy conditions, it will be useful to
denote $\hat{A}=A-(1/2)c(A)g,$ when $A$ is any second rank
covariant tensor. For instance, we observe easily that
$\hat{\hat{A}}=A,$ and $\hat{Ric}=E.$ We say $A$ is observer
non-negative definite if $A(u,u) \geq 0,$ for every time-like
vector $u,$ whereas we say that $A$ is dominantly non-negative if
$A(u,v) \geq 0$ whenever $u$ and $v$ are vectors with $g(u,v)<0.$
Picking a future half of the light cone arbitrarily at $m \in M,$
we note that the future light cone is an open subset of the
tangent space $T_mM,$ and therefore there is a basis for the
tangent space consisting of future time-like unit vectors, say
$u_1,u_2,u_3,u_4.$ In this frame we have $A_{\alpha \beta} \geq 0$
and $g_{\alpha \beta} < 0.$ If $\tilde{A}$ denotes the
transformation of $TM$ uniquely defined by
$g(\tilde{A}(u),v)=A(u,v)$ for all vectors $u,v$ over the same
base point, then saying $A$ is dominantly non-negative is
equivalent to requiring $-\tilde{A}(u)$ be future time-like or
null whenever $u$ is future time-like, which is the usual
statement of the dominant energy condition for $A.$  The {\it weak
energy condition} simply requires that $T$ is observer
non-negative definite, whereas the {\it strong energy condition}
requires that $\hat{T}$ is observer non-negative definite.  Thus,
by the Einstein equation, the strong energy condition is
equivalent to requiring that $Ric=8 \pi \hat{T}$ be observer
non-negative definite. As it is expected that the pressures and
stresses are smaller than the mass energy, it is reasonable that
$c(T) \leq 0,$ so the weak energy condition is probably weaker
than the strong energy condition. Notice the strong energy
condition is really completely geometric, as it says simply
$Ric(u,u) \geq 0$ for every time-like vector $u,$ and $Ric$ can be
determined by the connection without reference to the metric. We
say that the energy-stress tensor $T$ satisfies the {\it dominant
energy condition} provided that $T$ is dominantly non-negative.
Now, $\tilde{H}=2\tilde{T}-c(T)id_{TM}.$ Thus, the dominant energy
condition holds for $T,$ if and only if $-\tilde{H}(u)-c(T)u$ is
future time-like or null for any future time-like vector $u \in
TM.$ However, thinking of $Ric=4\pi H$ with $H$ the total energy,
it would now seem reasonable to require that $H$ and hence also
$Ric$ be dominantly non-negative.

\section{THE COSMOLOGICAL CONSTANT}

If we include the cosmological constant $\Lambda$ in the Einstein
equation, it becomes

\begin{equation}\label{cosmo1}
Ric -(1/2)Rg+\Lambda g=8\pi T,
\end{equation}
which is of course the same as

\begin{equation}\label{cosmo2}
Ric-(1/2)Rg=8\pi[T-(1/8\pi)\Lambda g],
\end{equation}
which means we view the equation here as having a modified
energy-stress tensor

\begin{equation}\label{cosmo3}
T_{\Lambda}=T-(1/8\pi)\Lambda g.
\end{equation}

We then have $c(T_{\Lambda})=c(T)-(1/2\pi)\Lambda,$ so the
effective energy-stress tensor of the gravitational field is

\begin{equation}\label{cosmo4}
T_g=T-c(T)g-(1/2\pi)\Lambda g,
\end{equation}
and the effective total energy-stress tensor serving as source is

\begin{equation}\label{cosmo5}
H=2T-c(T)g-(1/2\pi)\Lambda g.
\end{equation}
In any case, as $div~g=0,$ it follows that our conclusions about
the energy-momentum flow of the gravitational field from
(\ref{divgravenergy1}) and (\ref{divgravenergy2}) remain valid,
even in the presence of a cosmological constant.

\section{QUASI LOCAL MASS}

The problem of defining the energy contained in a space-like
hypersurface has led to many different definitions of the mass
enclosed by a closed space-like surface contained in an arbitrary
spacetime manifold, and these go by the general name quasi-local
mass. Typically, they are defined by some kind of surface integral
and give an indication of the mass enclosed by the space-like
surface. For an extensive survey of these we refer the interested
reader to \cite{SZ}. In particular, the results of \cite{TIPLER}
on the Penrose quasi-local mass show that the results can be
interesting when the space-like surface is not the boundary of a
space-like hypersurface.  A list of desirable properties of any
definition of quasi-local mass is given in \cite{YAU3}, where in
particular it is shown that for their definition, the quasi-local
mass enclosed by a space-like surface $S$ is non-negative provided
that the dominant energy condition holds and the surface $S$ is
the boundary of a hypersurface, $\Omega.$  It is further assumed
that the boundary surface $S$ has positive Gauss curvature and
space-like mean curvature vector, and consists of finitely many
connected components. The local energy condition assumed is framed
in terms of the second fundamental form of the hypersurface, and
in particular, we can see that for a geodesic hypersurface it
reduces to the condition that the scalar curvature of the
hypersurface, $\Omega,$ is non-negative, since in that case the
second fundamental form vanishes (extrinsic curvature zero). But,
the scalar curvature of the space-like hypersurface $\Omega$ is
$2E(u,u)=16 \pi T(u,u),$ where $u$ is a time-like future pointing
unit normal field on $\Omega.$  So if the energy-stress tensor
satisfies the weak energy condition in this case, then the energy
density as seen by observers riding the hypersurface is
non-negative, and we would simply integrate $(1/8 \pi)E(u,u)$ over
the hypersurface to find the energy inside, which is clearly
non-negative.  The amazing result in \cite{YAU3} is that the
quasi-local mass defined there is defined in terms of integrals
over the boundary $S.$  For instance, their results show if the
energy inside any one component of $S$ vanishes, then $S$ is
connected and $\Omega$ is flat (\cite{YAU3}, Theorem 1, page 183),
and thus the result shows that the energy in $\Omega$ is in some
sense determined by the geometry of the boundary and its mean
curvature vector under the assumptions stated above. In order to
make use of the total energy-stress tensor, $H,$ in a similar
setting, one would assume an appropriate energy condition, and
then for a space-like hypersurface $K$ with future time-like unit
normal field $u,$ it is natural to consider $H(u,u)\mu_K$ where
$\mu_K$ is the volume form due to the Riemannian metric induced on
$K.$ The integral of $H(u,u)\mu_K$ over all of $\Omega$ should be
the total energy inside $K.$ More generally, if we assume that $H$
is dominantly non-negative, that is, it satisfies the dominant
energy condition, then given another reference future pointing
time-like vector field $k,$ one might then integrate $H(u,k)
\mu_K$ over $K.$ If a 2-form $\alpha$ can be found on $K$
satisfying $d \alpha = H(u,k)\mu_K,$ and if $K$ is a 3-submanifold
with boundary $B,$ then by Stoke's theorem, the total energy
inside $K$ is the integral of $\alpha$ over the boundary $B$ of
$K.$ In particular, we say that $K$ is {\it instantaneously
static} if there is an open set $U \subset M$ containing $K$ and a
vector field $k$ on $U$ which is future pointing and orthogonal to
$K$ and which satisfies Killing's equation, at each point of $K.$
If $\omega=k^*$ is the dual 1-form to $k,$ so $\omega(v)=g(k,v)$
for all vectors $v,$ then this is equivalent to requiring
$Sym(\nabla \omega)|K=0$ or equivalently that $(d\omega)|K=2\nabla
\omega|K,$ which to be perfectly clear means that the difference
$d\omega-2\nabla \omega$ as calculated on $U$ in fact is zero at
each point of $K.$ Then as in the Komar \cite{KOMAR} integral (see
\cite{PINTONETOSOARES}, \cite{WALD}, pages 287-289 or
\cite{POISSON}, pages 149-151) it follows that

\begin{equation}\label{quasilocalmass0}
(-1/8 \pi)d*d\omega=(1/4 \pi)Ric(u,k)=H(u,k)\mu_K.
\end{equation}
Here, $*$ denotes the Hodge star operator on $M.$ Thus, $(-1/8
\pi)*d\omega$ is a potential for the total energy on $K.$ For any
closed 2-submanifold $S$ of $K$ we define the quasi-local total
energy $H(S,k)$ by

\begin{equation}\label{quasilocalmass1}
H(S,k)=-\frac{1}{8 \pi}\int_S *d(k^*).
\end{equation}
Thus, if $K_0 \subset K$ is a submanifold with boundary $S=\del
K_0,$ then by Stoke's Theorem, (\ref{quasilocalmass1}) becomes

\begin{equation}\label{quasilocalmass2}
H(S,k)=-\frac{1}{8 \pi}\int_{K_0} d*d(k^*)=\int_{K_0} H(u,k)
\mu_K,
\end{equation}
which is then non-negative if $H$ is observer non-negative
definite. Of course, this is the same as requiring $Ric$ be
observer non-negative definite, a purely geometric requirement.
Thus, if $H(S,k)=0,$ with $S=\del K_0,$ then by
(\ref{quasilocalmass2}), under the assumption that $H$ is
dominantly non-negative, we would conclude that $Ric(u,k)=0$ on
$K_0.$ But, this means that $Ric(u,u)=0$ on $K_0,$ and this means
that the scalar curvature of $K_0$ is identically zero. In
particular, if $K_0$ has constant sectional curvature, this would
imply that $K_0$ is actually flat. Notice that if we have an
asymptotically flat spacetime with a global time-like killing
vector field orthogonal to a spacelike slice, normalized to be a
unit vector at spatial infinity, then our definition of the
quasi-local total energy would be exactly the Komar mass which is
well known in the literature \cite{SZ}. Thus in the expression
$H(S,k),$ the normalization for $k$ is determined by requiring
that it be of unit length at the event at which the observer is
located.  If the observer is located so that $S$ is in the
observer's causal past, then it would seem we must assume that the
domain of $k$ contains this past light cone. In general, if $k$ is
a Killing field on all of the open set $U,$ then being orthogonal
to $K$ means (\cite{WALD}, page 119, (6.1.1)) that also $\omega
\wedge d\omega =0,$ where $\omega=k^*.$ Then (see \cite{WALD},
page 443, (C.3.12)) we find, using $f=ln(|g(k,k)|),$

\begin{equation}\label{quasilocalmass4}
d\omega=-\omega \wedge df,
\end{equation}
and using the fact that here $*[\omega \wedge df]=-(e^{f/2}D_nf)
\mu_S,$ where $n$ is the outward unit normal to $S=\del K_0,$ and
$\mu_S=dA$ is the area 2-form on $S,$ we obtain finally,

\begin{equation}\label{quasilocalmass5}
H(S,k)=-\frac{1}{8 \pi} \int_S e^{f/2}D_nf dA.
\end{equation}
In particular, for the vacuum Schwarzschild solution with mass
$M,$ taking the Killing field $k=\del_t,$ we see easily that the
mass calculated using the integral (\ref{quasilocalmass5}) gives
the value $M$ for the mass enclosed by any sphere centered at the
"origin" when we normalize the Killing field to be a unit vector
at infinity. On the other hand, if we calculate that value of the
integral by normalizing to make the Killing vector a unit at
radial coordinate $r_0,$ as $H(S,k)$ is homogeneous in $k,$ the
normalizing constant comes out resulting in

\begin{equation}\label{quasilocalmass3}
M_{r_0}=\frac{M}{[1-\frac{2M}{r_0}]^{1/2}}.
\end{equation}
Keeping in mind this is now the total energy, gravitational and
massive, this indicates a problem develops as $r_0 \rightarrow
2M,$ even though we know it is not a real problem for the
spacetime. This indicates the problem is probably due to the
normalization involving the Schwarzschild radial coordinate which
obviously breaks down at $r_0=2M.$ After all, what we are
integrating is equivalent by Stoke's Theorem to integrating
$H(u,k)\mu_K,$ when $S=\del K_0,$ and we really want to be
integrating $H(u,u)\mu_K.$ We do not have the actual potential.

\section{APPENDIX ON ANALYTIC CONTINUATION}

Suppose that $S:E \times E \lra F$ is any symmetric bilinear map
of vector spaces.  Let the quadratic function $f_S$ be defined by
$f_S(x)=S(x,x).$  Then $f_S$ determines $S.$  This is just (what
mathematicians would call) polar decomposition:

\begin{equation}\label{polardecomp}
S(x,y)=(1/4)[f_S(x+y)-f_S(x-y)].
\end{equation}
That is, we note that

\begin{equation}\label{polarpm}
f_S(x \pm y)=f_S(x)+f_S(y) \pm 2S(x,y),
\end{equation}
so subtracting the "minus" equation from the "plus" equation of
(\ref{polarpm}) gives (\ref{polardecomp}). In particular, if
$f_S=0,$ then $S=0.$ But more is true. For, suppose $E$ is a
topological vector space and that $f_S$ is constant on the open
subset $U \subset E.$  Choosing $x \in U$ and $y$ sufficiently
"small",we can assume that both $x+y$ and $x-y$ belong to $U$ in
which case we have $S(x,y)=0$ from the polar decomposition
(\ref{polardecomp}).  But from homogeneity, it follows that
$S(x,y)=0,$ for every $x \in U,$ and every $y \in E.$ Since $S$ is
symmetric, it follows that $S(y,x)=0,$ for every $x \in U,$ and
every $y \in E.$ Now, if $x,y \in E$ are any vectors, simply
choose a vector $x_0 \in U,$ and we have $S(x,x_0)=0,$ so
$S(x,y)=S(x,y-x_0),$ and we will see that $S(x,y-x_0)=0.$ For
there is a small scalar $t \neq 0$ so that $x_0+t(y-x_0)$ belongs
to $U,$ and therefore, $0=S(x,x_0+t(y-x_0))=S(x,x_0)+tS(x,y-x_0).$
Since $S(x,x_0)=0,$ it now follows that $S(x,y-x_0)=0,$ and thus
finally we have $S(x,y)=0.$ The argument can be simplified by
using differentiation (see for instance \cite{SACHSWU}, page 72),
but we prefer to give a purely algebraic argument here.  Using
differentiation, we next generalize easily to n-linear maps, but a
telescoping algebraic argument could be applied with a little more
effort.

Thus, if $S$ is a continuous symmetric $n-$linear map of the
Banach space $E$ into the Banach space $F,$ and if we define the
monomial function $f_S:E \lra F$ by the rule
$f_S(x)=S(x,x,x,...,x)=Sx^{(n)},$ then $f_S$ is an analytic
function. In fact, if $x_1,x_2,x_3,...,x_n \in E,$ then
differentiating we find

\begin{equation}\label{analyticcontinuation}
D_{x_1}D_{x_2}D_{x_3}...D_{x_n}f(a)=(n!)S(x_1,x_2,x_3,...,x_n),~a
\in E.
\end{equation}
From (\ref{analyticcontinuation}), we see very generally that if
$U$ is any open subset of $E$ on which $f_S$ is constant, then in
fact, $S=0,$ since we can choose $a \in U.$ Indeed, if $a \in U,$
since $f_S$ is constant on $U,$ it follows that the derivative on
the left side of the equation (\ref{analyticcontinuation}) is 0,
and hence also the right side, for every possible choice of
vectors $x_1,x_2,x_3,...x_n \in E.$ But notice that $a$ does not
appear on the right hand side of (\ref{analyticcontinuation}),
only $S(x_1,x_2,x_3,...x_n),$ and the vectors $x_1,x_2,x_3,...x_n$
can be chosen arbitrarily. Thus, $S=0$ follows. This is just a
very special case of the principle of analytic continuation.
However, to see (\ref{analyticcontinuation}), it served our
purpose for the Einstein equation to only examine the case where
$n=2$ and the vector spaces are finite dimensional, so the vector
topologies are unique and any bilinear map is therefore continuous
and in fact smooth. In this case, we have for any $x,w \in E,$ the
easily checked fact that

\begin{equation}\label{analyticcontinuation1}
D_wf_S(x)=S(x,w)+S(w,x)=2S(x,w).
\end{equation}
Differentiating again, since as a function of $x$ alone $D_wf_S$
is obviously linear, it follows that for any $x,v,w \in E,$

\begin{equation}\label{analyticcontinuation2}
D_vD_wf_S(x)=2S(v,w).
\end{equation}
It should be obvious how the general case of
(\ref{analyticcontinuation}) from the method of showing
(\ref{analyticcontinuation2}). In any case, clearly, if $f_S$ is
constant on some (no matter how small) non-empty open subset $U$
of $E,$ then choosing $x \in U$ would give $D_vD_wf_S(x)=0$ for
every $v,w \in E$ and every $x \in U.$ Thus the right hand side of
(\ref{analyticcontinuation2}) vanishes for every $v,w \in E$ and
this says $S=0.$ In our application, we are taking the vector
space to be the tangent space to $M$ at a specific point, say $m
\in M,$ so $E=T_mM.$ Then, either half of the light cone, future
or past, is an open subset of $T_mM$ and therefore, if $S$ is a
symmetric second rank tensor at $m$ whose quadratic form vanishes
on the future light cone or vanishes on the past light cone, or on
any small open subset of the light cone, then $S=0.$  It follows
that in general relativity theory, symmetric tensors at a point
can be specified by their monomial forms on the future or on the
past light cone at that point, and hence it suffices, by
homogeneity of the monomial forms to limit consideration to
time-like unit vectors. We might say in fact that in general
relativity, any tensor equation should be viewed as a tensor
valued symmetric tensor, and on all slots for which there is
symmetry one need only evaluate by using the same arbitrary unit
time-like vector in each of those slots. For instance, if
$A(a,b,c,d,e)$ is a $5-linear$ transformation which is symmetric
in the first three slots, then considering $A(a,b,c,\_,\_)$ as a
third rank symmetric tensor which is second rank tensor valued, it
is completely determined by knowing $A(u,u,u,\_,\_)$ for every
time-like unit vector $u \in TM.$ This is what we call the
symmetric tensor observer principle. In some sense, this is the
principle of relativity. If we apply this to the electromagnetic
field, $F,$ for instance, as it is anti-symmetric, that is, a
2-form, it should be viewed as a 1-form valued 1-form. Thus, it is
determined by giving the 1-form $F(u)$ for each time-like unit
vector $u.$ In fact, we can say that an observer with velocity $u$
holding a test charge $Q$ should feel the force $QF(u),$ so $F(u)$
is the force 1-form per unit charge experienced by an observer
holding an electric charge.  Since the force vector in the
instantaneous rest frame is orthogonal to the velocity, we must
have $F(u,u)=0$ here. As this experiment could be carried out by
any observer, this means that $F(u,u)=0$ for every time-like unit
vector, so $Sym(F)=0,$ by the symmetric tensor observer principle,
and therefore $F=Alt(F)$ must be a 2-form. We can similarly say
that any force field which applies to certain objects called
"charges", according to a scalar measure of that charge making
force felt by any observer holding the charge exactly proportional
to the charge, would be described by a 2-form.

\section{ACKNOWLEDGEMENTS}

I am very deeply indebted to Frank Tipler for many extremely
useful and helpful conversations and in particular, for making me
aware of the historic problems surrounding the energy density of
the gravitational field, as well as the recent developments in
quasi-local mass. I would also like to thank Toni Eastham and
Juliette Dupre for helpful conversations.

\med

\end{document}